\documentstyle[prl,aps]{revtex} 
\begin{document}
\begin{titlepage} 
\title{UHECR Production by a Compact Black-Hole Dynamo: Application to Sgr A*}
\author{Amir Levinson$^{1,2}$}
\address{School of Physics and Astronomy, Tel Aviv University, 
Tel Aviv 69978, Israel}
\author{ Elihu Boldt}
\address{Laboratory for High Energy Astrophysics, NASA 
Goddard Space Flight Center, Greenbelt, MD 20771, USA}
\maketitle  
\begin{abstract}
The possibility that the excess cosmic ray (CR) flux near 10$^{18}$ 
eV, reported
recently by the AGASA group, is due to a compact black hole dynamo
associated with the Sgr A* source is considered.  Under the assumption
that the Galactic center black hole rotates with nearly maximal spin,
and that the magnetic field threading the horizon is in rough equipartition
with matter accreted by the hole, the spectra and total fluxes
of accelerated CRs and their associated curvature emission depend only
on accretion rate.  For the accretion rate estimated on the basis of
observations of stellar winds near Sgr A* ($\sim 10^{-3}$ Eddington), the
maximum proton energy achievable by this mechanism is of the order of
$10^{18}$ eV, and the corresponding CR power
is $\sim 10^{41}\alpha_{CR}$ erg s$^{-1}$, where $\alpha_{CR}<<1$ is the CR
production efficiency.  The corresponding spectrum of curvature photons peaks
at around a hundred GeV, with a total luminosity comparable to the CR power
released.  For much lower accretion rates both the CR and gamma-ray fluxes
predicted are completely insignificant.  Upcoming gamma-ray experiments,
such as CANGAROO, HESS and GLAST, can be used to probe the parameters of this system.
\end{abstract}
\vspace{1in}
\noindent{Pacs Numbers: 98.70.Rz, 95.70.sa}\\
\noindent{Keywords: Cosmic rays; Gamma rays; Galactic center}\\

\vfill

$^1$ Corresponding author\\
$^2$ Email address:  Levinson@wise.tau.ac.il
\end{titlepage} 

\section{Introduction}
Recent analysis of data from AGASA \cite{Haetal99} and SUGAR 
\cite{Beetal00} have indicated an excess of cosmic ray intensity 
over a narrow energy range around
$10^{18}$ eV, that has been interpreted as due to a strong CR source in
the direction of the Galactic center.  The analysis by the AGASA group
suggested a rather extended source in the general vicinity of the Galactic
center, with the most significant excess obtained for a signal beam size
of about 20$^\circ$.  Close examination of the SUGAR data 
\cite{Beetal00}, that was
motivated by the results of the AGASA group, confirmed the existence of
an excess CR flux near the direction of the Galactic center, but indicated a
signal that appears to be consistent with a point source offset by about
7.5$^\circ$ from the true Galactic center.  Clay \cite{Cl00} suggested
that the pointlike excess is caused by neutrons produced as a result of
conversions of particles in a target located outside their
acceleration site; in order to reach Earth before decaying, the neutrons
should have energies in excess of $\sim10^{18}$ eV.
This naturally explains the absence of an excess at
energies below 10$^{18}$ eV.  The absence of an excess at energies well above
that is attributed to an upper limit on the acceleration energy in the source.
Clay (see also ref. \cite{Cletal00}) also proposed that the propagation of the
charged (non-converted) particles through the Galactic magnetic field 
might give rise to the more extended source that appears in the 
AGASA data.  Using detailed numerical simulations and reasonable models 
for the Galactic magnetic field he studied the
propagation of protons from the vicinity of the Galactic center to Earth, and
concluded that a source of protons near the Galactic
center would be unobservable at energies below $~10^{18}$ eV, but
should produce a halo of energy dependent size and location above this energy.
The details depend on the magnetic field model adopted, but in general
there should be a systematic shift of the source image to the north, owing
to the regular field component.

The above considerations motivated us to explore the application of a model,
developed recently to explain the origin of the highest energy cosmic rays
detected \cite{BG99}, to the Galactic center black
hole, Sgr A*.   In this model the ultra-high energy cosmic rays (UHECRs) are
accelerated by the electric potential difference generated
by spinning supermassive black holes associated with dormant AGNs.
The model assumes that those black holes were either formed with
critical angular momentum or spun up during
an earlier phase when the AGNs were active, and are presently rotating
with nearly maximal spins.  The basic picture envisaged is that a
small number of particles injected into the gap are being accelerated to
ultra-high energies during episodes when the {\em emf} induced by the 
hole dynamo
is not shorted out; in this picture the rotational energy of the hole is
liberated essentially in the form of UHE particles and their associated
curvature emission \cite{Le00}, rather than powerful jets,
as seen in blazars.  This is consistent with the expectation that spontaneous
breakdown of the vacuum should not occur in those systems \cite{Le00}.  For
the range of parameters relevant to UHECR production, the curvature losses
suffered by the accelerating particles are severe, suppressing the maximum
energy they can achieve substantially \cite{Le00,BL00}.  The
curvature photons are emitted predominantly in the TeV band, with an average
flux per UHECR source that exceeds the detection limit of current TeV 
experiments \cite{Le00}, and is conceivably highly variable.  The total 
power needed to be extracted from an average UHECR source in order to 
account for the measured CR flux above the GZK cutoff is rather small, less 
than about 0.1\% of the maximized Blandford-Znajek (BZ) power.  Why 
this process is so inefficient as compared with the process of jet 
formation in blazars is yet an open issue.  As argued by Punsley \& Coroniti 
\cite{PC90} the BZ power extracted from a BH should be governed by
the process of plasma injection in the ergosphere.  So perhaps in the 
UHECR sources, for which spontaneous vacuum breakdown as in blazars 
is not expected, plasma injection is considerably suppressed, resulting 
in a much lower efficiency.

Recent numerical simulations \cite{HK00} suggest that the accretion process 
and magnetic field structure in the vicinity of the horizon should be non 
stationary, owing to rapid magnetic field reconnection.  This would 
probably lead to appreciable complications of the (over) simplified 
model adopted in ref \cite{BG99}.
For instance the location of the gap and, perhaps, the voltage drop across
it might change with time, as well as the injection of seed particles 
into the gap.
How should this affect the picture described above is unclear at 
present.  Despite of all the complications anticipated, we shall 
proceed by using the simple model \cite{BG99} outlined above.
 
\section{Application to Sgr A*}
There are essentially three important parameters that determine the
CR and gamma-ray fluxes produced by the black hole dynamo: the mass of the
hole, its angular momentum, and the strength of the magnetic field threading
the horizon.  In the case of Sgr A* the mass is known to a reasonably
good accuracy from dynamical measurements.
The other two parameters, however, are highly uncertain.  Upper limits on the
power and spectral peak energy of emitted cosmic rays can be obtained if the
hole angular momentum is taken to be near its critical value, and the magnetic
field is assumed to be in equipartition with the matter accreted into 
the black hole.
Under these circumstances the results depend only on one parameter, namely the
accretion rate $\dot{m}$, henceforth measured in Eddington units 
($L_{Edd}/c^2$).
Below we adopt the estimate of Eckart \& Genzel \cite{EG97} for 
the black hole mass: $M=2.5\times10^6 M_{\odot}$.

The rate at which the Galactic black hole is accreting is 
unfortunately unknown.  The total luminosity associated with Sgr A* 
is of order $10^{37}$ erg s$^{-1}$ (see ref. \cite{Naetal98}).
If we naively relate this luminosity to accretion from a standard thin disk
with say 10 \% efficiency, we obtain $\dot{m}\sim10^{-7}$.  This 
value is well below those estimated by Melia \cite{Me92} 
($\dot{m}\sim10^{-1.5}$) and Genzel et al. \cite{Geetal94} 
($\dot{m}\sim10^{-3}$), based on observations of stellar winds in
the vicinity of Sgr A* and assuming Bondi accretion (that might be irrelevant
if the infalling matter possesses angular momentum).  Narayan et al. 
\cite{Naetal98} constructed an ADAF model for Sgr A* in an attempt to
reconcile the low luminosity of Sgr A* with the relatively high accretion
rates estimated by Melia and Genzel et al.  A viable fit of the spectrum of 
Sgr A* by the ADAF model (but cf. ref. \cite{Faetal98} for a different
interpretation) appears to be consistent with the measured mass of Sgr A*,
with an accretion rate of $\dot{m}\sim 10^{-3}$, and with an equipartition
magnetic field, provided the parameter $\delta$ is sufficiently small
($\delta\sim10^{-3}$), where $\delta$ represents the fraction of 
turbulent energy that is tapped for electron heating.  However,
it has been subsequently shown \cite{QN99} that models with large
$\delta$ and considerable mass loss rate via winds \cite{BB99},
for which only a small fraction of the inflowing matter ultimately reaches
the black hole, are also in agreement with observations.  Thus, it seems that
there is, at present, a large uncertainty in the value of $\dot{m}$, but that
it is likely to lie in the range $10^{-7}<\dot{m}<10^{-3}$.

The strength of the equipartition magnetic field depends, quite generally, on
the details of the inflow pattern in the vicinity of the horizon.  A 
rough estimate gives $B_4\simeq 61 (\dot{m}/M_6)^{1/2}$, where 
$M_6=M/10^6M_{\odot}$ \cite{BG99}.
This is in good agreement with the mean equipartition field 
calculated using the self-similar ADAF model by Narayan and Yi \cite{NY95}.  
For Sgr A* ($M_6=2.5$) we obtain
\begin{equation}
B_4 \simeq 38 \dot{m}^{1/2}.
\label{B}
\end{equation}

The {\em emf} generated by a rotating black hole is given approximately
by $\Delta V\sim
4.5\times10^{17}(a/M)(h/R_g)^2 B_4M_6$ V.  Taking $a=M$, $h=R_g$, $M_6=2.5$,
and using eq. (\ref{B}) yields
\begin{equation}
\Delta V\simeq5\times10^{19} \dot{m}^{1/2}\ \ \ {\rm V}.
\end{equation}
For the range of $\dot{m}$ considered here ($10^{-7}$ to $10^{-3}$) the
{\em emf} given by the last equation lies in the range between about
$10^{16}$ and $1.5\times10^{18}$ Volt.

The energy change per unit length of an accelerating particle of charge $Z$ and
energy $\epsilon=\gamma m_ic^2$ can be expressed as \cite{Le00},
\begin{equation}
\frac{d\epsilon}{ds}=\frac{eZ\Delta V}{h} - 
\frac{2}{3}\frac{e^2Z^2\gamma^4}{\rho^2},
\label{dE}
\end{equation}
where $\rho$ is the mean curvature radius of magnetic field lines in the gap.
The first term on the r.h.s of the last equation accounts for the energy gain
owing to the acceleration by the gap electric field, and the second 
term accounts
for the losses due to curvature radiation.
Under the assumption that $\rho$ is independent of the particle's energy,
eq. (\ref{dE}) can be solved analytically.  The maximum energy achievable,
$\epsilon_{max}$, can then be obtained by integrating eq. (\ref{dE}) over the
gap (i.e., from $s=0$ to $s=h$), and is given implicitly by,
\begin{equation}
\ln\frac{1+x}{1-x} + 2\tan^{-1}(x)= 4\eta^{-1}.
\label{x}
\end{equation}
Here $\eta=3 \mu 
(a/M)^{-3/4}M_6^{-1/2}Z^{-5/4}B_4^{-3/4}(\rho/R_g)^{1/2}(h/R_g)^{-7/4}$,
with $\mu$ being the mass of the ion in units of the proton mass,
is the suppression factor defined in eq. (5) of ref. \cite{Le00} for $a=M$, and
$x=\epsilon_{max}/\bar{\epsilon}_{max}$, where $\bar{\epsilon}_{max}$, 
given explicitly in eq. (4) of ref. \cite{Le00}, is the maximum 
acceleration energy in the limit of large
suppression (i.e., strong curvature losses), as can be readily seen from eq.
(\ref{x}) wherefore $x\sim 1$ when $\eta<<1$.  In the opposite limit, 
$\eta>>1$,
the solution to eq. (\ref{x}) is
given to a good approximation by $\epsilon_{max}\simeq eZ\Delta V$, 
as expected.
For the parameters adopted above we find
\begin{equation}
\eta\simeq 0.12\mu Z^{-5/4}\dot{m}^{-3/8},
\label{eta}
\end{equation}
where it has been assumed that $\rho=h=R_g$.

The maximum rotational power that can be extracted from a Kerr black hole
is given roughly by $L_{BH}\sim 10^{40}(a/M)^2M_6^2B_4^2$ erg s$^{-1}$.
We suppose that a fraction $\alpha_{CR}<<1$ of this power is liberated in the
form of CRs.  Taking again $a=M$, and using eq. (\ref{B}), we find that the
CR power released by the Galactic BH can be written as
\begin{equation}
L_{CR}\sim 10^{44} \alpha_{CR}\dot{m} \ \ \ {\rm erg \ s^{-1}}.
\label{Lcr}
\end{equation}

The relative production efficiency of curvature photons, defined as 
the ratio of
the total curvature loss per nucleus and the maximum energy gain,
can now be expressed as
\begin{equation}
\kappa\equiv \frac{(Ph/c)}{\epsilon_{max}}\simeq 0.5 Z^5B_4^3\mu^{-4}
(\epsilon_{max}/eZ\Delta V)^3
\simeq 8\times10^4 Z^5\mu^{-4}\dot{m}^{3/2}(\epsilon_{max}/eZ\Delta V)^3,
\label{delta}
\end{equation}
where the dimensionless parameter, $\epsilon_{max}/eZ\Delta V$, can be computed
by employing eqs. (\ref{x}) and (\ref{eta}) once $\dot{m}$ is specified.
The limit of large suppression corresponds to $\kappa>>1$,
whereas $\kappa<<1$ corresponds to small curvature losses.
To a good approximation then the luminosity associated with
curvature emission is given by $L_{\gamma}\simeq\kappa L_{CR}$.
The curvature spectrum will peak at an energy \cite{Le00}
\begin{equation}
\epsilon_{\gamma}\sim60(Z/\mu)^3B_4^3 = 3\times10^6(Z/\mu)^3\dot{m}^{3/2}
\ \ \ {\rm GeV}.
\label{Eg}
\end{equation}

We now use the results derived above to estimate the luminosity and
spectral peak energy of emitted CRs and curvature photons, for
different accretion rates corresponding to different models of Sgr A*.
Consider first the possibility of standard accretion with a very low rate.
Adopting $\dot{m}=10^{-7}$ (see above), eq. (\ref{eta}) yields 
$\eta\simeq 51$ for
protons and $\eta\simeq 47$ for iron.  Substituting the latter result into
eq. (\ref{x}), one finds $\epsilon_{max}\simeq 1.5\times10^{16}$ eV 
for protons and
$\epsilon_{max}\simeq 4\times10^{17}$ eV for iron.  The spectrum of curvature
photons should peak at around 100 keV if the accelerated particles are
predominantly protons and 10 KeV if iron.
From eq. (\ref{Lcr}) we obtain a CR power of
$L_{CR}\simeq10^{37} \alpha_{CR}$ erg s$^{-1}$, and from eq. (\ref{delta})
we obtain a radiative efficiency of $\kappa\simeq 10^{-5.5}$
for both protons and iron and a corresponding gamma-ray luminosity of
$L_{\gamma}\simeq 3\times10^{31}\alpha_{CR}$ erg s$^{-1}$.  We 
therefore conclude
that for such a low accretion rate the fluxes of CRs and curvature photons
produced by the dynamo mechanism are completely negligible.

We consider next the situation whereby the radio - through X-ray
spectrum of Sgr A* is
produced by an ADAF.  Adopting the accretion rate obtained from the
standard ADAF model \cite{Naetal98}, viz., $\dot{m}\simeq 
10^{-3}$, and
repeating the above calculations we find: $\epsilon_{max}\simeq 10^{18}$
eV for protons and
$\epsilon_{max}\simeq 2.5\times10^{19}$ eV for iron, a CR power of
$L_{CR}\simeq10^{41} \alpha_{CR}$ erg s$^{-1}$, a radiative efficiency of
$\kappa\simeq 1$ for both protons and iron with a corresponding
gamma-ray luminosity $L_{\gamma}\simeq L_{CR} $, and peak energy
for the curvature spectrum of approximately 100 GeV or 10 GeV, depending on
whether the accelerated particles are, respectively, protons or iron.
We stress that the estimate of the peak energy is highly uncertain because
of the strong dependence of $\epsilon_{\gamma}$ on the strength of the
magnetic field (see eq. [\ref{Eg}]).
Taking for illustration $\alpha_{CR}=10^{-3}$, which is roughly the value
inferred for the dormant AGNs \cite{BG99,Le00}, yields $L_{CR}\sim
L_{\gamma}\sim 10^{38}$ erg s$^{-1}$. The corresponding gamma-ray flux
at Earth ($S_{\gamma}\sim 10^{-8}$ erg cm$^{-2}$ s$^{-1}$) is several
orders of magnitude above the threshold sensitivity
anticipated for the HESS \cite{Ko99} and CANGAROO \cite{Taetal99} imaging 
atmospheric Cerenkov telescopes in the southern
hemisphere.  The photon detections to be obtained with GLAST
during a normal one-year scan mode are expected to be on the order of a
thousand counts or more in this case (Thompson 2000, personal communication).

In addition to the energy losses corresponding to curvature
radiation we must also address the possible impact of inelastic pair
and pion producing collisions with the environmental electromagnetic
radiation near the hole (see ref. \cite{BL00}). We
estimate that, during the acceleration of
a proton to $10^{18}$ eV by the putative dynamo associated with Sgr A*,
there should be negligible drag arising from inelastic pair producing
collisions with ambient photons.  In particular, since the Lorentz
gamma factor is never larger than $10^9$ during the acceleration
process, only submillimeter (wavelength $<$ 1.2 mm) ambient photons are
involved in electron-positron pair production.  From the estimated
SED of Sgr A* \cite{Faetal98,BD99} we conclude
that the corresponding number density of suitable  submillimeter
photons implies a radiation length about two orders of magnitude
larger than the intrinsic size of the radio source \cite{Loetal99}.
Finally, it is important to note that collisions involving pion
generation are of substantially higher inelasticity than those
involving pair production \cite{St68} and thereby potentially
much more catastrophic for the putative
dynamo action. However, for inelastic collisions with photo-pion
production, only those ambient photons of wavelength shorter than 8
microns are involved; the associated collision mean-free-path in this
instance is then estimated to be even larger than the radiation
length characterizing pair production.

\section{Discussion}
In this paper we have considered the application of a compact black
hole dynamo mechanism to CR production by Sgr A*.  We have argued that
if the rate at which matter is accreted by the black hole is on the order
of that estimated on the basis of observations of stellar winds near
Sgr A*, then protons of maximum energy $\sim10^{18}$ eV and total
power $\sim 10^{38}$ erg s$^{-1}$ (assuming CR production efficiency
of $10^{-3}$) can be produced
by the dynamo mechanism, provided the black hole angular momentum is near
its critical value, and assuming equipartition between the magnetic field
threading the horizon and inflowing matter.  An associated gamma-ray emission
arising from curvature acceleration, with a luminosity comparable to the
total CR power released, and with a spectrum that peaks sharply at around
100 GeV is predicted under these conditions, and motivates observations
with future missions.  Positive detection of curvature photons will
have important implications for the process of accretion into the black
hole.

Whether the above scenario can explain the excess CR flux reported by
the AGASA group recently, even in the case of large accretion rate, remains to
be checked; at these energies propagation effects appear to be quite
sensitive to the Galactic magnetic field model adopted \cite{Cl00}
and should be accounted for properly.  The energy flux of the apparent
point source offset from the galactic center observed with SUGAR
is $3\times10^{-11}$ ergs s$^{-1}$ cm$^{-2}$ \cite{Beetal00}.  Assuming
its distance from Earth to be comparable to that of Sgr A*, yields
a luminosity of $\sim 2\times10^{35}$ ergs/(s cm$^2$).  If the SUGAR source
is associated with
energetic neutrons produced in a target there by the conversion of
UHECR protons (or Fe nuclei) originating from Sgr A*, then given the
error circle of SUGAR, it implies that the target size should lie in the
range between 30 to 300 pc.

Finally, it is interesting to note that the galactic center region is 
in fact a known gamma-ray source \cite{Metal98},
observed over the band 30 MeV to 10 GeV with the EGRET instrument on
the Compton GRO to exhibit a hard power law spectrum that steepens
above  2 GeV and where the inferred energy flux within the total
measured band is  $~3\times10^{-9}$ ergs s$^{-1}$ cm$^{-2}$; however, 
the emission
volume is likely extended, and the portion arising from a possibly
major point source is still uncertain (an issue that should be much
better addressed with the improved angular resolution to be provided
by GLAST).  If we naively associate the break with the peak energy of curvature
photons, we obtain from eq. (\ref{Eg}) $\dot{m}\simeq 10^{-4.5}$.  From
eqs. (\ref{Lcr}) and (\ref{delta}) we obtain a luminosity of $L_{\gamma}
\sim 10^{38}\alpha_{CR}$ erg s$^{-1}$.  Assuming that about half the flux
is emitted by the point source \cite{Metal98}, we 
obtain a CR production efficiency of $\alpha_{CR}< 0.1$, which seems rather 
high compared with the efficiency inferred for the dormant AGNs.  However,
since $\epsilon_{\gamma}\propto B^3$, as seen from eq. (\ref{Eg}), even
relatively small deviations from equipartition would lead to considerably
lower peak energy and, consequently, higher values of
$\dot{m}$ and lower values of $\alpha_{CR}$.  Alternatively,
the potential curvature radiation from Sgr A*  might well be a
separate gamma ray component from that detectable with EGRET (e.g.,
at a much higher energy, such as characteristic of the case
illustrated here).
 
We thank Michael Loewenstein, Gustavo Medina-Tanko and David Thompson
for enlightening discussions.

\end{document}